\documentstyle{article}
\begin{document}
\begin{center}
{\Large \bf Heron Variables in 3-body Coulomb Problem }\footnote{
The updated version of the talk published in the Proceedings of 
the XI International  Workshop on High Energy Physics and Quantum Field
Theory, 12-18 September 1996, St.-Petersburg, Russia. pp. 403-405. Ed. by
B.B.Levtchenko, Moscow, 1997.}
\\ \vspace{4mm}

V.S.Vanyashin\footnote{e-mail: vvanyash@ff.dsu.dp.ua,
vanyashv@heron.itep.ru}\\
Dnepropetrovsk State University, 320625, Dnepropetrovsk \\
Ukraine\\
\end{center}

\begin{abstract}
The use of coordinate variables with independent physical
boundaries -- Heron variables -- is proposed for the 3-body
problem. The ansatz is given for variational trial wave
functions without local energy infinities at the Coulomb
singularities. 
\end{abstract}

\section*{Introduction. Heron variables}

The wave function of a bound system in a ground state
depends only on interparticle distances. There are three
distances $r_1, r_2$ and $r_3$ for 3-body systems, six
for 4-body and so on. Though the distances are locally
independent and admit unconstrained partial
differentiation, the physical boundaries for each of them
depend on the values of the others. The triangle
inequalities take place:
\begin{equation}\begin{array}{rcl}
 0\,\, \le & | r_2 - r_3 |\,\, \le & r_1 \,\, \le \,\,r_2 + r_3 \,\,  <
 \,\, \infty, \\
 0\,\, \le & | r_1 - r_3 |\,\, \le & r_2\,\,  \le\,\, r_1 + r_3 \,\,  <
 \,\,\infty, \\
 0\,\, \le & | r_1 - r_2 |\,\, \le & r_3\,\,  \le\,\, r_1 + r_2 \,\,  <
\,\,\infty.  \end{array}                                      \label{eq:1}
\end{equation}

More symmetrical Hylleraas variables~\cite{Hylleraas}
\begin{equation}
          s = r_1 + r_2, \,\, t = | r_1 - r_2 |,\,\, u = r_3       \label{eq:2}
\end{equation}
are used for decades in the variational calculations of the
3-body systems. These variables are also subjected to the
physical region inequalities:

\begin{equation}
  0 \le t \le u \le s < \infty  .                      \label{eq:3}
\end{equation}

All such variables with mutual constraints are not
fully independent, and this causes definite technical
difficulties in calculations.

Meanwhile, there are coordinate variables with
independent physical boundaries. They are known for
millenniums since the discovery by Heron (Alexandria,
1st century A.D. ) of the triangle area formula~\cite{Heron}

\begin{equation}
  S = \sqrt{p(p-a)(p-b)(p-c)},\,\, p=(a+b+c)/2 .         \label{eq:4}
\end{equation}

The Heron variables
\begin{equation}\begin{array}{rl}

      p - a & = h_1 = (-r_1 + r_2 + r_3)/2, \\
      p - b & = h_2 = ( r_1 - r_2 + r_3)/2, \\
      p - c & = h_3 = ( r_1 + r_2 - r_3)/2
\end{array}                                          \label{eq:5}
\end{equation}
span separately over the physical region

\begin{equation}
     0 \le h_1 < \infty,\,\,  0 \le h_2 < \infty,\,\, 0 \le h_3 < \infty.
\label{eq:6}\end{equation}

They are fully independent, that leads to great
simplification in  variational calculations. The repeated
integrals turn to multiple integrals,  which reduce to
one-dimensional integrals, if the trial wave  functions
are appropriately chosen.

The equivalent variables, without referring to Heron and now usually named
as perimetric variables,
were proposed in~\cite{Coolidge} and effectively used in~\cite{Pekeris}
and other works.\footnote{The author is indebted to G. W. F. Drake for
these references.}

\section{Ansatz for variational trial wave functions}

For Helium and Helium-like ions the simplest ground
state function can be  chosen as follows:

\begin{equation}
      \psi  = \exp ( -Z r_1 -Z r_2 + r_3/2 ).           \label{eq:7}
\end{equation}

If an electron pair were isolated, the exact solution
$\psi_{ee} = \exp(r_3/2)$ of the Schr$\ddot{\mbox{o}}$dinger equation with
repulsive Coulomb potential would be unphysical because of
unrestricted exponential growth. Being embedded into the
wave function of a whole bound system, the exponent with
the plus sign in front of $r_3$ does not produce
unrestricted exponential growth in the 3-body
configuration space. Transition to Heron variables
\begin{equation}
    r_1 = h_2 + h_3,\,\, r_2 = h_1 + h_3,\,\, r_3 = h_1 + h_2  \label{eq:8}
\end{equation}
exposes exponential decrease on each variable:
\begin{equation}
    \psi = \exp (-(Z-1/2)h_1-(Z-1/2)h_2-2 Z h_3).        \label{eq:9}
\end{equation}

The mean energy value, calculated with this function is

\begin{equation}
   H_{mean} = -{(-1 + 2 Z)(1 - 8 Z + 28 Z^2 - 64 Z^3)
   \over {4(-1 + 10 Z - 32 Z^2)}}.
                                                   \label{eq:10}
\end{equation}

This value is obtained without  any parameters for
adjustment, as at this lowest level of approximation no
minimization procedure was performed.

It is instructive to compare the approximation qualities of the
proposed wave function with that of the variational wave
function with one adjustable parameter -- the effective
charge:

\begin{equation}
      Z_{eff} = Z - 5/16,\,\,\,\, \psi_{eff} = \exp (-Z_{eff} r_1 -Z_{eff}
r_2).                                                 \label{eq:11}
\end{equation}

The one parameter mean energy value

\begin{equation}
        H_{mean}(Z_{eff})  = -( Z - 5/16)^2         \label{eq:12}
\end{equation}

shows for all $Z$  only  a little bit bigger deviation from the energy
eigenvalues, as the parameterless formula~(\ref{eq:10}) does, but the local
properties of $\psi_{eff}$ are much worse than that of $\psi$.

The so-called local energy, defined as

\begin{equation}
      E_{local}  = H \psi_{eff}  /   \psi_{eff}       ,
                                                           \label{eq:13}
\end{equation}
takes infinite values at the Coulomb
singularities~\cite{Bethe}.

On the contrary, the local energy, attributed to the
proposed function~(\ref{eq:7}), has bounded variation in the whole
physical region of variables.

As one can judge with the known literature, almost all
used up to now variational wave functions are plagued
with the local energy infinities, in spite of the huge
number of their parameters.

How to avoid local energy infinities at any
approximation stage? The simplest ansatz for good trial
function is suggested on by the presented example of the
wave function in the lowest approximation. It is
sufficient to have a linear superposition of triple
products, formed by the exact Coulomb solutions for each
pair of the particles involved.

The building blocks are:
\begin{equation}\begin{array}{rl}
    \psi (n, l=0) & = \exp(-a r/n)\, _1F_1 (1 - n, 2, 2a r/n), \\
  \psi_i (n, l=1) & = r_i\, \exp(-a r/n)\, _1F_1 (2 - n, 4, 2ar/n), \\
 \psi_{ij} (n, l=2) & =(\delta_{ij} r^2 - 3 r_i r_j)\,\exp(-a r/n)\,
 _1F_1(3-n,6,2a r/n),\\ \cdots  \,\,\,\,\,\,  .
\end{array}                                          \label{eq:14}
\end{equation}

In (~\ref{eq:14}) $\vec r = \vec r_1, \vec r_2$, or $\vec r_3$, $n = n_1,
n_2$, or $n_3$ and $a = Z$, or $a = -1/2$ correspondingly.

The individual principal quantum number $n$ need not to be
integer, and can be treated as an adjustable parameter in
a truncated hypergeometrical series. Each term of the linear
superposition should not violate the condition of
integral convergence: $Max(n_1, n_2)\, <\, 2 Z n_3$.
The intermediate vector indices should be summed in all possible ways
according to transformation properties of the whole system
state. And the symmetrization or antisymmetrization on the variables $\vec
r_1,\,\vec r_2 $ should be performed.

\section{Axial states of a 3-body system}

Let's apply the proposed ansatz to the 3-body state, when
the relative angular momenta of each pair of particles form an overall
$P$-state with positive parity -- the axial state.

In the lowest approximation the axial wave function is

\begin{equation}
    \tilde \psi_i =
\exp (-Z r_1/2 -Z r_2/2+r_3/4)\epsilon_{ijk} r_{1j} r_{2k} .
\label{eq:15} \end{equation}

The mean energy value is calculated as

\begin{equation}
         {H^{ax}}_{mean}  =
{(-1 + 2 Z)(-1 + 14 Z - 84 Z^2 + 280 Z^3 - 512 Z^4) \over
 {16(-1 + 16 Z - 100 Z^2 + 256 Z^3)}}.
\label{eq:16}\end{equation}

Though the energy level of this doubly excited state
is well above the ground level, the axial state  
is stable against autoionization. For symmetry reasons, the autoionization
of an axial state cannot proceed to the final 2-body ground state, which
is scalar state. For energy reasons, it also cannot proceed to the final
2-body $2P$-state, that follows from the lowest approximation formula
(~\ref{eq:16}) for $Z \, > \, 1$ and was established for the Hydrogen
negative ion in~\cite{Drake}.

\section*{Acknowledgements}

 The author would like to thank V.I.Savrin and E.E.Boos
for the invitation to the XI International Workshop on
High Energy Physics and Quantum Field Theory and A.Yu.Voronin
for discussion and valuable remarks.

\end{document}